\documentclass{iopart}

\usepackage{graphicx}
\begin{document}
\title{Tsunamis, Viscosity and the HBT Puzzle}
\author{Scott Pratt}
\address{Department of Physics and Astronomy,
Michigan State University\\
East Lansing, Michigan 48824-1321, USA}
\date{\today}

\begin{abstract}
The equation of state and bulk and shear viscosities are shown to be able to affect the transverse dynamics of a central heavy ion collision. The net entropy, along with the femtoscopic radii are shown to be affected at the 10-20\% level by both shear and bulk viscosity. The degree to which these effects help build a tsunami-like pulse is also discussed.
\end{abstract} 

\maketitle

Two-particle correlation measurements provide a strict six-dimensional test of heavy-ion collision dynamics. The experimental correlation function, $C({\bf P},{\bf q})$, can be used to determine the source function, $S({\bf P},{\bf r})$, which describes the probability that two particles of equal velocity, whose total momentum is ${\bf P}$, are separated by ${\bf r}$ in their asymptotic trajectory. Several sophisticated transport models (mainly hydrodynamic models with soft equations of state) have grossly failed to describe correlations. Several microsopic models, that have effectively stiffer equations of state, have come closer to the data, but still have difficulty at the 10-20\% level. In particular, they tend to underpredict $R_{\rm side}$, the sideward dimension of the source function, relative to $R_{\rm long}$, the longitudinal dimension along the beam axis, and $R_{\rm out}$, the outward dimension which is parallel to ${\bf P}$ if one has boosted longitudinally to a frame where the longitudinal component of ${\bf P}$ vanishes.

The difficulties mentioned above have been branded as the ``HBT puzzle'' (Here HBT refers to Hanbury-Brown and Twiss, the founders of a similar technique that was successful in determining stellar diameters). Using this puzzle as motivation, this talk will provide a survey of how changes in the dynamics, notably changing the equation of state or introducing bulk or shear viscosity, affect correlation measurements. The chief findings of this survey are that viscous effects can change the shape of the source function at the 10-20\% level, and in a direction that may make hydrodynamic models more consistent with data.

The most direct way in which the equation of state manifests itself in HBT measurements is through the entropy, which can be determined in an approximately model-independent fashion from correlations and spectra. The extraction relies on the relation between entropy and the phase space density, $f({\bf p},{\bf r},t)$,
\begin{eqnarray}
S &= (2J+1)\int \frac{d^3r d^3p}{(2\pi)^3} \left[ -f\ln f \pm (1\pm f)\ln (1\pm f) \right]
\label{entropy}\\
\nonumber
&\approx (2J+1)\int \frac{d^3r d^3p}{(2\pi)^3} \left[ -f\ln f + f \pm f^2/2\right],
\end{eqnarray} 

The phase space density can be expressed in terms of the Gaussian fit to the source distribution,
\begin{equation}
f({\bf p},{\bf r},t)=\frac{\sqrt{\lambda}}{(2\pi R_{\rm inv}({\bf p})^2)^{3/2}}
\exp\left\{-\frac{({\bf r}-{\bf R}_{\bf p}(t))^2}{2R_{\rm inv}^2({\bf p})}\right\}.
\end{equation}
Here, $\sqrt{\lambda}$ is the fraction of particles not emitted from long-lived resonances -- as the entropy from the long-lived resonances will be accounted for separately. The center of the phase space cloud, ${\bf R}_{\bf p}$, does not affect the answer. Thus, the inference of the entropy is model-independent aside from the assumption of a Gaussian profile in coordinate space.

Values of $R_{\rm inv}({\bf p})$ and spectra were compiled from data for a variety of particles, using data from the 130$A$ GeV Au+Au run at RHIC. Whereas some of the radii were taken from correlations anlayses, others were taken from coalescence ratios. For instance, the ratio of $\phi$ spectra to the square of the kaon spectra were used to determine $R_{\rm inv}({\bf p})$ for kaons. For some particles, such as hyperons, proton radii were substituted. This lead to an estimate, $dS/dy=4450\pm 10\%$ \cite{subrata}. This is not a surprising answer given that the number of hadrons at mid-rapidity approached one thousand, that the entropy per pion would be expected to be between three and four units, and that the entropy per particle for baryons could be twice that of pions.

If the analysis were repeated using the greatly improved data from the 200$A$ GeV Au+Au runs during the last few years, the uncertainty could be significantly reduced. In particular, much of the uncertainty of the previous estimate derived from uncertainties in spectra. Uncertainties in the radii were less important since the enter logarithmically. Thus, if all the volumes were misidentified by a factor of two, the entropy would be mis-evaluated by $\ln(2)=0.69$ units per particle, or 15\%. A more careful analysis should be able to reduce the uncertainties to approximately 5\%. 

Assuming that entropy is conserved from early times, e.g., one fm/c, where the volume is known, a measurement of entropy represents an equation of state if it can be compared to an estimate of the initial energy density. Unfortunately, unlike the entropy per unit rapidity $dS/dy$, $dE/dy$ is not conserved due to the longitudinal work done during the expansion, which requires a model-dependent estimate. Even if entropy is produced during the expansion, an upper bound on the initial entropy is established by the analysis.

\begin{figure}
\centerline{\includegraphics[width=0.5\textwidth]{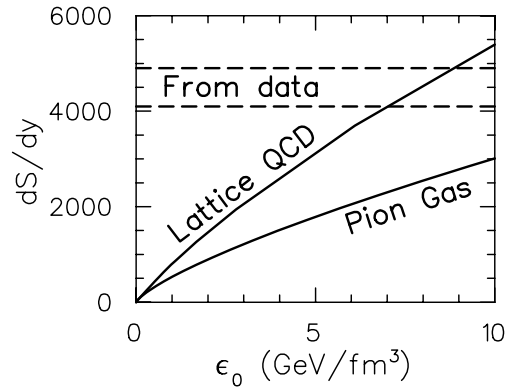}}
\caption{\label{fig:entropy}
The entropy per unit rapidity extracted from \cite{subrata} of $\sim$4450 units is compared to lattice predictions, which depend on the unknown initial energy density. Lattice results are consistent with the extraction if average energy density at $\tau=1$ fm/$c$ is in the neighborhood of 8 GeV/fm$^3$. However, it needs to be emphasized that this comparison ignores entropy created during the expansion.}
\end{figure}

Figure \ref{fig:entropy} displays $dS/dy$ as a function of the initial energy density $\epsilon$ for $\tau=1$ fm/c from lattice gauge theory which provides the entropy density $s$ as a function of $\epsilon$. The mapping required an assumption that the volume for one unit of rapidity at a time $\tau=1$ fm/$c$ is $4\pi R^2\tau$, i.e., $dS/dy=4\pi R^2\tau s$. The Bjorken estimate of the initial energy density uses the transverse energy, and ignores the longitudinal motion of the particles at breakup, and ignores the longitudinal work performed by the expansion. Thus, the lattice value for $dS/dy$ is obscured by the uncertainty in the initial energy density which might be anywhere between 6 and 10 GeV/fm$^3$. Nonetheless, it is remarkable to see how the lattice expectation fits well into the experimental constraint. However, one should remain cognizant of the fact that entropy is essentially a logarithmic quantity, and a 10\% agreement is equivalent to explaining the volumes at the 50\% level.

The entropy analysis shows that if a transport model fits spectra but systematically overpredicts or underpredicts the HBT volume, the product $R_{\rm out}R_{\rm side}R_{\rm long}$, will similarly fail to produce the entropy. Furthermore, the models should be able to adjust the HBT volume by adjusting the equation of state.

The shape of the source distribution provides additional information, which is crucial for understanding the expansion dynamics. The lifetime of the collision will become most manifest in $R_{\rm long}$, which, for a high-energy collision with a large rapidity extent, depends on the thermal velocity and velocity gradient,
\begin{equation}
R_{\rm long}\sim\frac{v_{\rm therm}}{dv/dz}.
\end{equation}
This relation ensues from the fact that the phase space  cloud for particles of zero rapidity are constrained in the longitudinal dimension by the thermal velocity being able to overcome the collective longitudinal velocity, which increases as $z\cdot dv/dz$. If one neglects longitudinal acceleration, which is a valid approximation when the rapidity spread is high, the velocity gradient becomes $dv/dz=1/\tau$. Hence, $R_{\rm long}=v_{\rm therm}\tau$, and given an estimate of the thermal velocity, one can determine the lifetime $\tau$. Including longitudinal acceleration in the analysis should increase the estimated lifetime, at the 10\% level or less \cite{prattanalhydro}.

The outward and sideward dimensions tend to be similar if the matter disintegrates instantaneously. However, surface emission allows some particles to leave earlier, and gain a ``head start'' which results in extended values of $R_{\rm out}$ relative to $R_{\rm side}$. As the HBT puzzle involves finding a reason to extend $R_{\rm side}$ relative to $R_{\rm out}$ and $R_{\rm long}$, one could imagine several modifications to the dynamics that could make a model more consistent with data:
\begin{enumerate}
\item Increasing early acceleration \cite{teaney,heinz,romatschke}. This might be accomplished by shear effects which increase the transverse components of the stress-energy tensor at early times. This is natural from both the perspectives of viscous hydrodynamics and of classical longitudinal fields. For a classical electric field pointing in the $z$ direction, the stress energy tensor becomes $T_{xx}=T_{yy}=\epsilon, T_{zz}=-\epsilon$. This provides three times the effective pressure in the transverse direction as that of an ultrarelativistic gas. The negative effective pressure in the $z$ direction describes the fact that pulling capacitor plates apart requires work, in contrast to a positive pressure system which does work when it expands.
\item Supercooling \cite{csorgocsernaisupercooling}. A super-cooled system might, by virtue of the lower pressure at the outer border of the fireball, have reduced surface emission and be more likely to dissolve suddenly. This possibility has not yet been pursued in a quantitative model. However, the related phenomena of bulk viscosity in the region of $T_c$ will be presented further below.
\item Surface emission \cite{heiselberg}. A simple explanation of the effect involves confining emission to the surface. However, given that the energy from the center must at some time escape, and that pions with transverse momenta more than 100 MeV/$c$ move faster than the surface of the fireball, increasing the component from surface emission typically leads to the opposite behavior and increases $R_{\rm out}/R_{\rm side}$. However, the development of a strong pulse, could effectively enhance the density of the surface relative to the center, and lead to enhanced emission from the surface while emitting the particles in a sudden burst. This possibility will be described below and likened to a tsunami.
\item Refraction by the mean field \cite{cramermiller,prattrefraction}. After pions leave the point of their last interaction, their trajectories can be bent by the mean field. Attractive mean fields indeed lead to increased relative values for $R_{\rm side}$, and have been proposed as a solution to the HBT puzzle. However, such fields must be remarkably strong, for breakup densities. The Cramer-Miller paper assumed a scalar field that would reduce the masses of the pions to 50 MeV/$c^2$, but a virial expansion in terms of the densities suggests the strength should be an order of magnitude smaller \cite{prattvirial}.
\end{enumerate}

In this paper I will concentrate on the effect of adding bulk and shear viscosity, and changing the equation of state. The studies are based on a simple hydrodynamic model, which assumes cylindrical symmetry and boost invariance, thus effectively reducing the problem to one dimension. Numerical solutions are collected by employing a co-moving mesh that maintains local simultaneity by having the time step vary with position \cite{prattanalhydro}. Thus, each neighboring mesh point refers to space-time points which are simultaneous according to a co-moving observer. In some ways this simplifies the numerics. Both shear and bulk viscosity are incorporated through the Israel-Stewart form, \cite{muronga,heinz,romatschke}. The treatment varies somewhat from previous efforts, but details will be provided later in a more technical paper. The equation of state employed will be that of a simple hadronic gas for temperatures below $T_c=165$ MeV, or equivalently for energy densities below the maximum hadronic energy density, $\epsilon_{\rm max}^{(h)}$. Only the standard baryon decuplet and octet, and meson spin 0 and 1 meson flavor octets and singlets are included. Once the energy density exceeds the maximum allowed for a hadron gas, the equation of state assumes a constant speed of sound, $c_{\rm s,mixed}$, until the energy density has increased by an amount $L$. If $c_{\rm s,mixed}$ were zero, it would correspond to a first-order phase transition with a latent heat per volume $L$. Thus, the equation of state can be softened by either increasing $L$ or by decreasing $c_{\rm s,mixed}$. Once the energy density exceeds that of the ``mixed'' phase, $L+\epsilon_{\rm max}^{(h)}$, the speed of sound is set to a higher value, $c_s^2=0.3$, consistent with lattice results at high temperature.

Calculations were begun at an early proper time, 0.25 fm/$c$, with an energy density profile adjusted to maintain a fixed transverse energy per unit rapidity. Calculations were run for a low energy density, with a breakup hyper-surface chosen by the outermost point with energy density above 50 MeV/fm$^3$. Particle emission was calculated with Cooper-Frye criteria. Rather than carefully calculating correlation functions from the emission points of pions and generating effective Gaussian radii, which would only be consistent with experiment, r.m.s. radii are generated from direct pions, i.e., those that are not the product of decays. The breakup criteria and HBT radii estimates are certainly crude and not state of the art, but should be sufficient for illustrating the importance of the equation of state and viscosity.

\begin{figure}
\centerline{\includegraphics[width=0.4\textwidth]{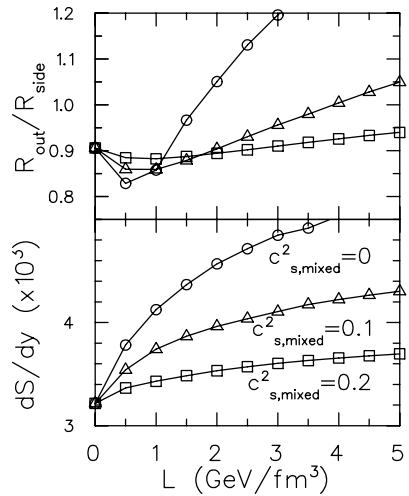}}
\caption{\label{fig:hbt_varyc2}
Lower Panel: The entropy density is shown for three different speeds of sound during the ``mixed'' phase, as a function of the width of the phase. For $c_{\rm s,mixed}=0$, the transition is first order and $L$ is the latent heat. For softer equation (larger $L$ and lower $c_s^2$) larger entropies ensue.\\
Upper Panel: The $R_{\rm out}/R_{\rm side}$ ratio is shown for the same value of $c_{\rm s,mixed}$. Due to the crude breakup criteria and the neglect of decay pions, the ratio should somewhat underestimate the experimental one, but the trends should be robust. It appears difficult to simultaneous match the experimental entropy density and have a small $R_{\rm out}/R_{\rm side}$ ratio.}
\end{figure}
Figure \ref{fig:hbt_varyc2} shows how the equation of state affects the net entropy per unit rapidity, and the source size measurements. A fairly soft equation of state is required to match the measured entropy density, $\sim 4450$. The experimental $R_{\rm out}/R_{\rm side}$ ratio is near unity, which given the neglect of decay pions here, should be under-predicted in the upper panel. The degree to which the inclusion of decays will alter these results is not known. However, the sensitivity to the equation of state is clear. 

\begin{figure}
\centerline{\includegraphics[width=0.45\textwidth]{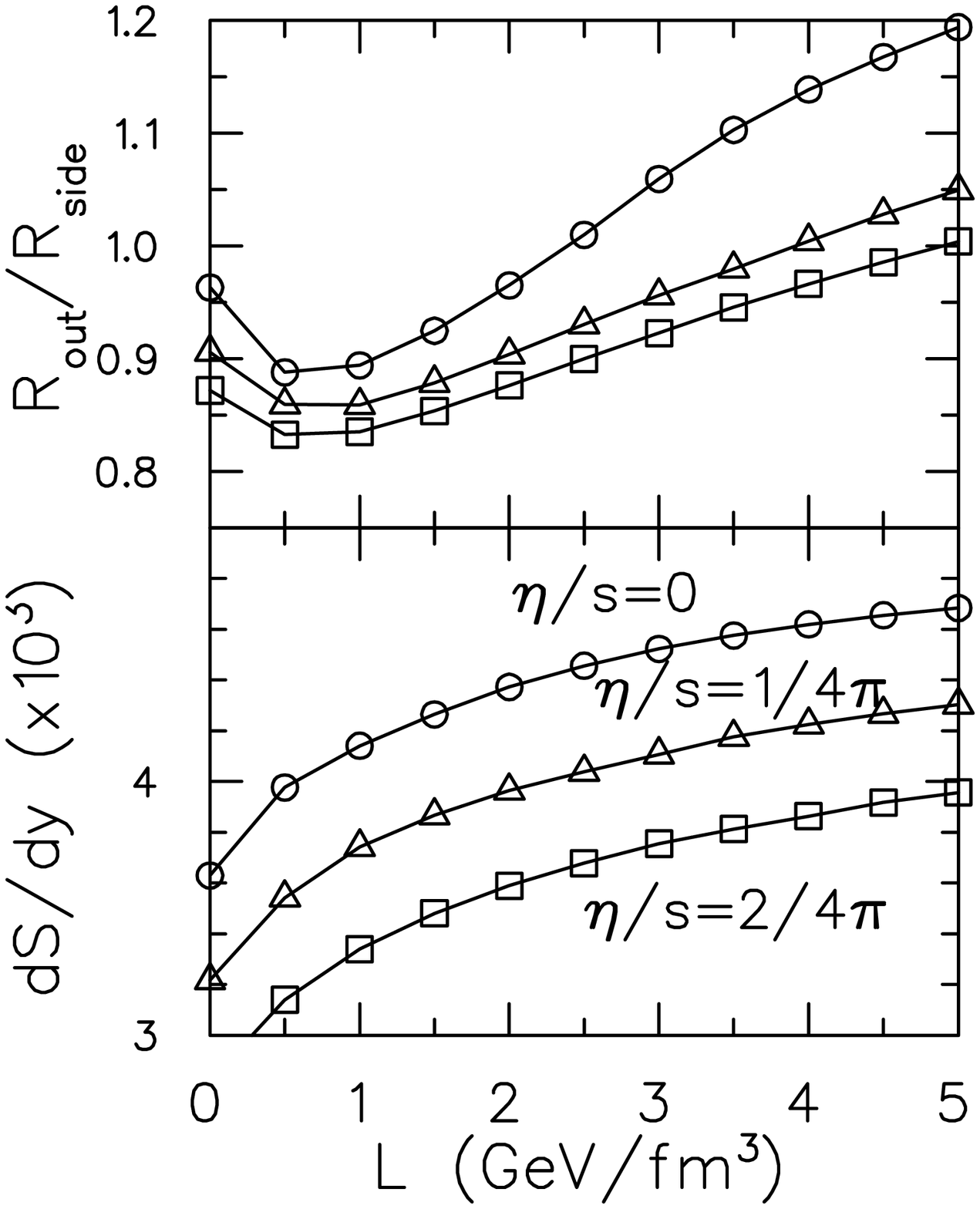}~~~
\includegraphics[width=0.45\textwidth]{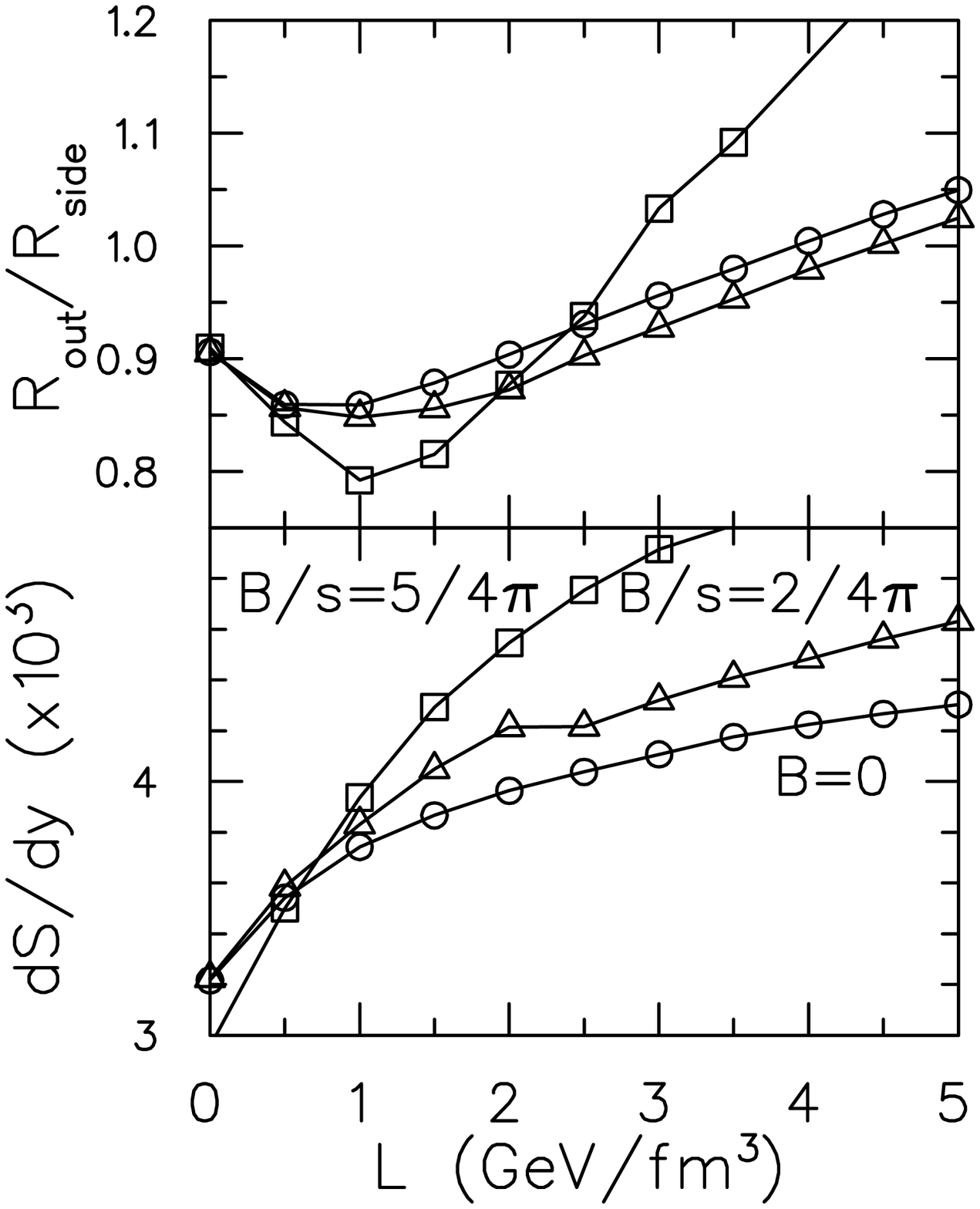}}
\caption{\label{fig:hbt_varyetaB}
Left-Hand Side: The same as Fig. \ref{fig:hbt_varyc2}, only that the speed of sound in the mixed phase is fixed at 0.1, and the viscosity is varied, $4\pi\eta/s=0,1,2$. The higher viscosity lead to lower entropy densities and lower $R_{\rm out}/R_{\rm side}$ ratios.\\
Right-Hand Side: Fixing the shear viscosity at $\eta/s=1/4\pi$, and the speed of sound in the ``mixed'' phase at 0.1, the entropy density and $R_{\rm out}/R_{\rm side}$ ratios are shown for three different values of the bulk viscosity. Since the bulk viscosity is set to zero outside the ``mixed phase'', the effects are stronger for larger $L$. Adding a bulk viscosity seems to simultaneously accommodate larger entropy densities and smaller $R_{\rm out}/R_{\rm side}$ ratios.
}
\end{figure}

The left-side panels of Fig. \ref{fig:hbt_varyetaB} illustrate the sensitivity of the results to shear viscosity. For these calculations the speed of sound in the mixed phase was fixed at $c_{\rm s,mixed}^2=0.1c$, and the bulk viscosity was set to zero. Somewhat counter-intuitively, the entropy density was reduced by incorporating stronger shear. This is due to the increased transverse acceleration, which increases the ratio of collective to thermal energy. Given that the initial profile was adjusted to fix the final transverse energy, higher shear resulted in lower entropy despite the fact that viscous effects create entropy. 

The effects of bulk viscosity are shown in Fig. \ref{fig:hbt_varyetaB}. In these calculations, the speed of sound in the ``mixed'' phase was fixed at 0.1, and the shear viscosity was fixed at $\eta/s=1/4\pi$. Although one expects the bulk viscosity to be small for temperatures far above $T_c$, large effects are expected near $T_c$, which should be associated with the difficulty with maintaining equilibrium in a region where the equilibrium values of the mean fields are rapidly changing \cite{paech,kharzeev}. For these calculations the bulk viscosity was chosen to be zero outside the mixed-phase region. For energy densities between the boundaries of the mixed phase, the bulk viscosity was set to a maximum of either $B_{\rm max}/s=2/4\pi$ or $5/4\pi$. The maximum value was applied to the midpoint, then linearly varied to zero at the boundaries of the mixed phase. For many of the calculations described in Fig. \ref{fig:hbt_varyetaB}, the effective pressure, $T_{xx}+T_{yy}+T_{zz}$, became negative due to the bulk viscosity. For such strong effects, one might want to address non-equilibrium dynamics directly. For instance, one might simultaneously solve equations of motion for the mean $\sigma$ field alongside those for viscous hydrodynamics as was done in \cite{dumitru}.

Unlike the shear viscosity, larger bulk viscosities led to larger entropy densities. The $R_{\rm out}/R_{\rm side}$ ratios also increased with larger bulk viscosities for large $L$, but decrease for small values of $L$. One feature of the results is that a large bulk viscosity allows one to simultaneously explain both a larger entropy density and a small $R_{\rm out}/R_{\rm side}$ ratio. Although quantitative conclusions are precluded by the sloppy treatment of the breakup stage, it is clear that bulk viscosity has the potential to significantly alter the conclusions of femtoscopic analyses.

\begin{figure}
\centerline{\includegraphics[width=0.5\textwidth]{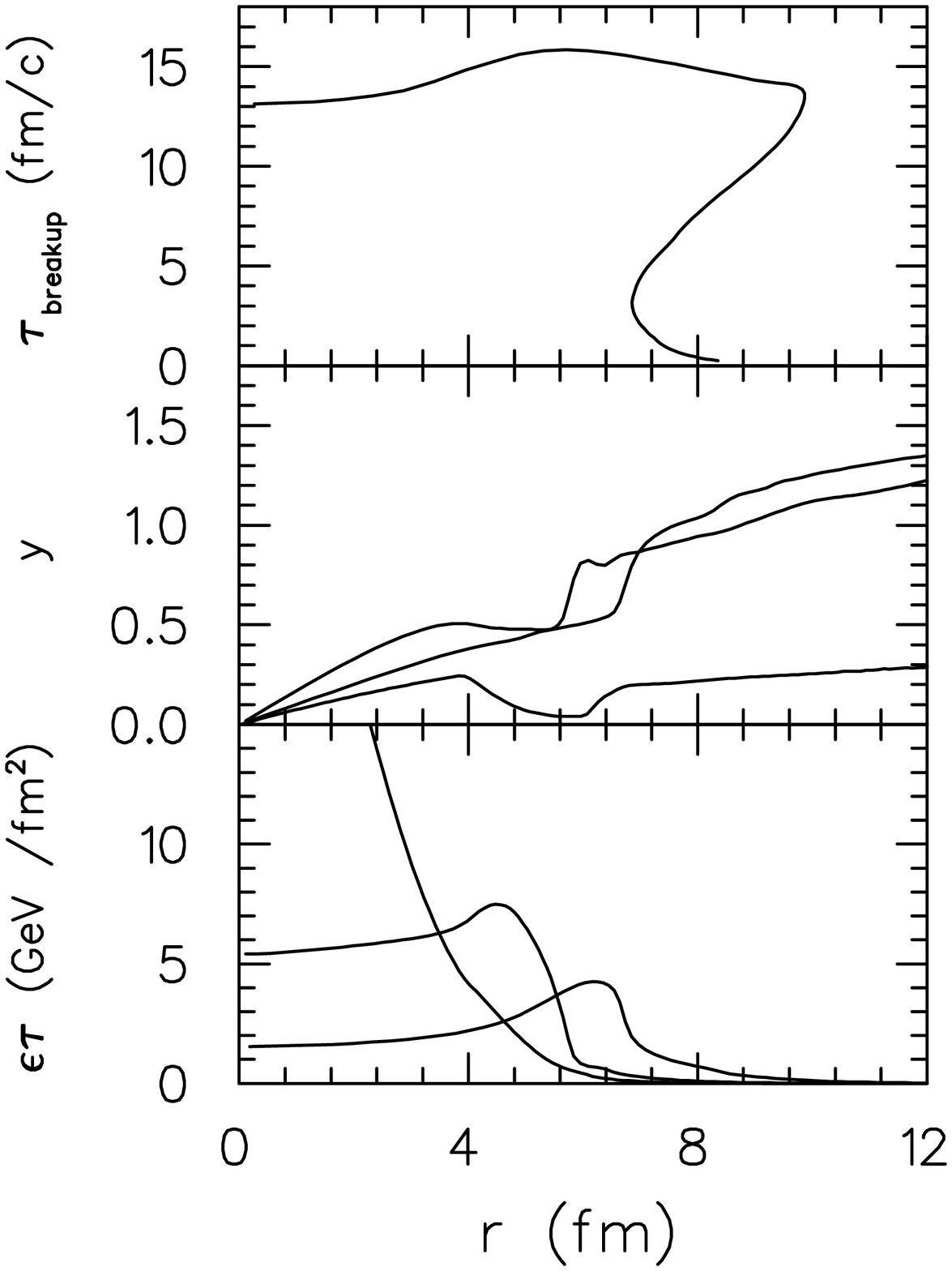}}
\caption{\label{fig:profile}
The profile of the energy density (multiplied by the proper time) for a soft equation of state is shown in the lower panel for three times, $\tau=1,5,9$ fm/$c$. A clear pulse develops due to the matter pushing into the mixed-phase region, where the matter ceases to accelerate as the speed of sound goes to zero. The transverse rapidity profile is given in the middle panel, which highlights the region where the velocity gradients conspire to build the pulse. The breakup hyper surface is displayed in the upper panel, and shows how the matter tends to break up suddenly after an extended time over which emission is confined to the surface. Pulses can lead to a shortened $R_{\rm out}/R_{\rm side}$ ratio.
}
\end{figure}
As a final topic, I discuss the tsunami-like pulses that can form during the transverse expansion. A tsunami is generated when a traveling wave moves into a region with a smaller speed of sound (For a surface wave, the speed is $\sqrt{gh}$ where $h$ is the depth of the water). The amplitude then grows inversely proportional to the shrinking wavelength. For RHIC collisions, the matter pushes into the ``mixed'' phase region, where the speed of sound is smaller than the inner region. For central collisions at RHIC, where the interior energy density is initially several times higher than that of the mixed-phase region, the pulse is stronger for softer equations of state. Figure \ref{fig:profile} displays the transverse blast profile for such a rather extreme example, with $L=4$ GeV/fm$^3$, and zero speed of sound in the mixed phase. A motivation for investigating the pulse is to understand how it might affect the $R_{\rm out}/R_{\rm side}$ ratio. Indeed, the profile of the pulse will lead to a shorter ratio, but this trend is overwhelmed by the fact that building a strong pulse requires a soft equation of state, which by virtue of its higher entropy, requires a longer time to burn. In fact, in the example illustrated in Fig. \ref{fig:profile}, approximately a third of the particles are emitted from the surface before the sudden dissolution of the bulk part of the collision volume. A less dramatic pulse can be produced from the effects of bulk viscosity with a less soft equation of state, where, as illustrated in Fig. \ref{fig:hbt_varyetaB}, the $R_{\rm out}/R_{\rm side}$ ratio can be quite small while simultaneously leading to a large entropy.

The investigations reported on here are far from complete, and can only serve as a lesson concerning the dangers of interpreting experimental results, especially correlations, using models that neglect viscosity. As seen here, the effects of even a minimal shear viscosity change correlations and entropy at the level of tens of percent. The effects of bulk viscosity are similar in magnitude. However, the potential range of bulk viscosities are less well understood. For this reason, a determined effort to extract bulk viscosity from lattice data, perhaps along the lines proposed in \cite{kharzeev}, would be extremely welcome.

\section*{Acknowledgments}
Support was provided by the U.S. Department of Energy, Grant No. DE-FG02-03ER41259.

\end{document}